\documentstyle[12pt,epsfig]{article}

\newcommand{\eq}[1]{eq.(\ref{#1})}
\newcommand{\dpar}[2]{\frac{\partial #1}{\partial #2}}

\def\const{\mbox{const}}
\def\e{\mbox{e}}

\begin{document}
\title{Long-ranged forces and energy non-conservation in (1+1)-dimensions}
\author{
   V.~A.~Rubakov\\
    {\small \em Institute for Nuclear Research of the Russian Academy of
  Sciences,}\\
  {\small \em 60th October Anniversary prospect 7a, Moscow 117312}\\
  }

\maketitle
\begin{abstract}
We consider whether local and causal non-conservation of energy
may occur in generally covariant
theories with long-ranged fields (analogs of Newton's gravity)
whose source is energy--momentum. We find that such a possibility exists
in (1+1) dimensions.
\end{abstract}

\vskip .5 in


\section{Introduction}
Generally covariant (1+1)-dimensional theories provide convenient 
framework for considering various suspected properties of quantum gravity
in (3+1) dimensions (for reviews see, e.g., 
refs.\cite{Ginsparg,StromingerReview}).
A special feature of (1+1)-dimensional models 
is that some of them admit the
interpretation as string theories in higher dimensional target space.

The simplest model of this sort is literally the theory of closed 
strings in  Minkowski target space of critical 
dimensions. Indeed,
macroscopic  string states may be interpreted as 
(1+1)-dimensional  universes \cite{Hawking2,LyHa,VR,Nirov}.
However, one important feature present in (3+1)-dimensional gravity 
is missing in the simplest stringy model. Namely, in (3+1) dimensions there
exists long-ranged gravitational field whose source in energy--momentum
(Newton's gravity law), while there is no such field in that stringy model.

A particularly simple (1+1)-dimensional model where the mass (energy--momentum
of matter fields) produces long-range effects, is the dilaton gravity 
with matter
that has been widely discussed from the point of view of black hole
physics \cite{CGHS} (for careful analysis of the notion of
ADM mass in that model see refs.\cite{BilalKogan,Bilal}). Here we 
do not consider black holes, 
so we
simplify the model as much as possible. In particular, we set
the (1+1)-dimensional cosmological
constant to zero. 

An issue that we discuss in this paper within the framework of
(1+1)-dimensional dilaton gravity is whether energy conservation
is absolute necessity in models exhibiting  Newton's gravity law
or its analogs.
Intuitively, one may suspect that the existence of the long-ranged 
field associated with energy and momentum may be an obstacle to energy 
non-conservation in local processes. Indeed, naively any local process 
in which energy changes would lead to the 
instantaneous change of the long-ranged
field everywhere in space. This ``action-at-a-distance'' would itself
contradict locality and causality.
We will see, however, that this is not the case, at least in (1+1) 
dimensions.

A motivation for our analysis comes from the long-lasting discussion of
baby universes/wormholes and their possible effects on the parent universe
\cite{Hawking1,LRT,GS1,Coleman1,GS2,Banks}. It has been argued 
\cite{VR,VRPRD} that the emission of baby universes in some
(1+1)-dimensional models, including dilaton gravity, necessarily leads to
energy non-conservation in the parent universe. This in fact may be the case
in most theories allowing for baby universes/wormholes: these processes may 
give rise to the loss of quantum coherence in the parent universe, and
it has been argued on general grounds \cite{BanksPS}
that the energy non-conservation is inevitable in modifications of
quantum mechanics allowing for the loss of quantum coherence (see,
however, refs.\cite{EllisEnergy,Unruh}). 

We will not go into physics of baby universes/wormholes in this paper,
and take wore ``phenomenological'' point of view. Namely, we will assume 
that classical physics is valid everywhere in space-time except for its
small region where unknown processes may violate field equations and
 lead to non-conservation of energy. The question we address is
whether the latter is compatible with locality and causality (no
action-at-a-distance).

\section{Model and classical solutions}

The action for the simplest version of (1+1)-dimensional
dilaton gravity with conformal matter can be written in the form
similar to ref.\cite{Russo},
\begin{equation}
  S= -\frac{1}{\pi} \int~d^2\sigma~ \sqrt{-g}\left(-\frac{\gamma^2}{4}\phi R
          +g^{\alpha\beta}\partial_{\alpha}f^i \partial_{\beta}f^i \right)
\label{8+}
\end{equation}
where $\phi$ is the dilaton field, $f^i$ are matter fields,
and $\gamma$ is a positive coupling constant analogous to the Planck mass
of (3+1)-dimensional gravity. The coupling constant $\gamma$ may be absorbed
into the dilaton field, but we will not do this for book-keeping purposes.
The field equations are simplified in the
conformal gauge
\[
       g_{\alpha\beta} = \e^{2\rho}\eta_{\alpha\beta}
\]
where $\eta$ is the Minkowskian metrics in (1+1) dimensions. In this gauge,
the fields $\rho$, $\phi$ and $f^i$ obey massless free field equations.
There are also constraints
\begin{equation}
   -\frac{1}{2}\gamma^2\left(\partial_{\pm}\phi \partial_{\pm}\rho
         - \frac{1}{2}\partial_{\pm}^2 \phi\right) +
    \frac{1}{2} \left(\partial_{\pm}{\bf f}\right)^2 = 0
\label{8*}
\end{equation}
ensuring that the total energy--momentum tensor vanishes.

Let us outline some classical solutions in this model. We consider 
infinite one-dimensional space, $\sigma^1 \in (-\infty,+\infty)$,
and study localized distributions of matter. In this case one can 
further specify the gauge and choose
\begin{equation}
  \rho=0
\label{9b}
\end{equation}
so that the space-time is flat. 
Let us make one point here. Unlike the conformal gauge choice, the
further specification of the gauge, eq.(\ref{9b}), is possible only if
the field equations are satisfied everywhere in space-time. Indeed,
the residual gauge transformations in the conformal gauge
are
\[
    \sigma_{+} \to  \sigma_{+}' =   \sigma_{+}'(\sigma_{+})\;,~~~~~
    \sigma_{-} \to  \sigma_{-}' =   \sigma_{-}'(\sigma_{-})
\]
These can be used to set $\rho = 0$ only if $\rho$ obeys the massless
free field equation $\partial_{+}\partial_{-}\rho =0$ (whose general
solution is $\rho = \rho_{+}(\sigma_{+}) + \rho_{-}(\sigma_{-})$).

Once the gauge (\ref{9b}) is chosen,
equation (\ref{8*}) determines
 the dilaton field $\phi$ for a given matter distribution. Indeed, 
the solution to \eq{8*} is, up to an arbitrary linear function of coordinates,
\[
   \phi(\sigma) = \phi_{+}(\sigma_{+}) +
                   \phi_{-}(\sigma_{-}) 
\]
with
\begin{equation}
   \phi_{\pm} = -\frac{1}{\gamma^2} \int~d\sigma_{\pm}'~
                |\sigma_{\pm} - \sigma'_{\pm}|~
                (\partial_{\pm}{\bf f})^2 (\sigma'_{\pm})
\label{9a}
\end{equation}
Hence, the energy--momentum of matter fields produces long-ranged dilaton 
field which has linear  behavior at large $|\sigma^1|$. In
particular, the ADM mass can be defined as follows,
\begin{equation}
      \mu_{ADM} = - \frac{\gamma^2}{2\pi}
              \left[\dpar{\phi}{\sigma^1}(\sigma^1 \to +\infty) -
               \dpar{\phi}{\sigma^1}(\sigma^1 \to -\infty) \right]
\label{9c}
\end{equation}
In virtue of \eq{9a} it is equal to
\[
     \mu_{ADM} = \int_{-\infty}^{+\infty}~d\sigma^1~\varepsilon_M(\sigma)
\]
where
\[
    \varepsilon_M = \frac{1}{2\pi}\left[ (\partial_0 {\bf f})^2
                       + (\partial_1 {\bf f})^2 \right]
\]
is the energy density of matter.

It is instructive 
to study two narrow pulses of matter moving left and right and colliding at
$\sigma^1 = 0$. These pulses may be
approximated by the delta-function distribution,
\begin{equation}
     (\partial_{\pm} {\bf f})^2 = \frac{\pi}{2} \mu \delta (\sigma_{\pm})
\label{13a}
\end{equation}
where the normalization is such that the constant $\mu$  coincides with the
ADM mass.
In this case \eq{9a} has particularly simple form 
\begin{equation}
    \phi_{\pm} = 
                 - \frac{\pi}{2\gamma^2} (\mu |\sigma_{\pm}| 
                    + C\sigma_{\pm})
\label{13b}
\end{equation}
where an arbitrary integration constant is chosen to be the same for
$\phi_{+}$ and $\phi_{-}$ .

The dilaton field produced by two lumps of matter of equal energy and 
opposite momenta, moving towards each other (or from each other) with the speed
of light, is shown in fig.1. Needless to say, the linear dependence 
of $\phi$ on $\sigma^1$ at large $|\sigma^1|$ is nothing but the Coulomb
behavior of long-ranged field in one-dimensional space. In this respect
the dilaton field in (1+1) dimensions is analogous to gravitational field
of Newton's law in (3+1) dimensions.

\begin{figure}[htb]
\unitlength=0.60mm
\special{em:linewidth 0.4pt}
\linethickness{0.4pt}
\begin{picture}(150.00,148.00)
\put(80.00,106.00){\vector(0,1){34.00}}
\bezier{312}(90.00,109.00)(94.00,148.00)(98.00,109.00)
\bezier{28}(98.00,109.00)(98.00,106.00)(102.00,106.00)
\bezier{28}(90.00,109.00)(89.00,106.00)(85.00,106.00)
\bezier{312}(62.00,109.00)(66.00,148.00)(70.00,109.00)
\bezier{28}(70.00,109.00)(70.00,106.00)(74.00,106.00)
\bezier{28}(62.00,109.00)(61.00,106.00)(57.00,106.00)
\put(85.00,137.00){\makebox(0,0)[cc]{$\varepsilon_M$}}
\put(150.00,101.00){\makebox(0,0)[cc]{$\sigma^1$}}
\put(80.00,27.00){\vector(0,1){70.00}}
\put(69.00,72.00){\line(1,0){22.00}}
\put(150.00,79.00){\makebox(0,0)[cc]{$\sigma^1$}}
\put(83.00,95.00){\makebox(0,0)[cc]{$\phi$}}
\put(10.00,85.00){\vector(1,0){140.00}}
\put(10.00,106.00){\vector(1,0){140.00}}
\put(99.00,68.00){\line(6,-5){43.00}}
\put(140.00,32.33){\line(0,0){0.00}}
\put(61.00,68.00){\line(-6,-5){43.00}}
\bezier{36}(61.00,68.00)(66.00,72.00)(69.00,72.00)
\bezier{48}(89.00,72.00)(95.00,72.00)(99.00,68.00)
\end{picture}
\centerline{Fig.~1.}
\end{figure}

\section{Energy non-conservation}
Until now we have considered conventional classical theory. Let us now assume
that in a smalll region of space-time the field equations do not hold,
and the energy of matter is not conserved.

At first sight, there appears to be a conflict between 
the locality of the 
non-conservation of matter energy, and the existence, in the gauge 
(\ref{9b}), of the long-ranged dilaton field whose strength is
determined by the matter energy. Let us see that 
this conflict is only apparent.

Let us again consider the collision of two narrow pulses of matter.
Let us assume that the non-conservation of energy occurs when and where the
two pulses collide; let us also assume that the total momentum
does not change in the center of mass frame of the colliding pulses.
If we were able to use the gauge (\ref{9b}) everywhere in space-time,
we would have to introduce action-at-a-distance, as the field $\phi$
after the collision of the two pulses would be given by the
expression similar to eq.(\ref{13b}) but with matter distribution
characterized by the final ADM mass $\mu_f$ which is different from the
initial ADM mass $\mu_i$. However, we cannot insist on imposing the
gauge (\ref{9b}) everywhere in space-time, 
as the field $\rho$  
 does not obey the field 
equation $\partial_{\alpha} \partial^{\alpha} \rho =0$
in the collision region at the collision time, by our assumption.
We can choose the gauge $\rho =0$ for the initial configuration only.

If we insist on locality (no action-at-a-distance) we have to require
that the field $\phi$ has the same asymptotics as
eq.(\ref{13b}) with $\mu=\mu_i$ even {\it after} the collision. Then it is 
possible to set this field equal to eq.(\ref{9a}), with
${\bf f}$ equal to the {\it initial} field distribution,
everywhere in space-time. In other words, for infinitely narrow 
pulses we set $\phi$ equal to
\begin{equation}
    \phi_{i,\pm} = \phi_{f,\pm} = 
                 - \frac{\pi}{2\gamma^2} (\mu_i |\sigma_{\pm}| 
                    + C\sigma_{\pm})
\label{13bb}
\end{equation}
both before and after the collision. Hereafter the subscripts $i$ and $f$
refer to the fields before and after the collision.

The requirement (\ref{13bb}) of course implies a certain gauge choice
for the field configuration after the collision. The field $\rho$
in the final state is no longer zero, and has to be found by solving 
eq.(\ref{8*}) with the final distribution of matter. More precisely, we
have to find $\rho_f$ from the following equations,
\begin{equation}
   -\frac{1}{2}\gamma^2\partial_{\pm}\phi_i \partial_{\pm}\rho_f
         + \frac{1}{4}\gamma^2\partial_{\pm}^2 \phi_i +
    \frac{1}{2} \left(\partial_{\pm}{\bf f}_f\right)^2 = 0\;.
\label{8*b}
\end{equation}
Here $\phi_i$ is given by eq.(\ref{13bb}). We obtain in the case of 
infinitely narrow pulses
\begin{equation}
\rho_f = -\frac{1}{2C} (\mu_i - \mu_f) [\epsilon (\sigma_{+}) +
            \epsilon (\sigma_{-}) ]
\label{rho}
\end{equation}
where $\epsilon$ is the usual step function. The integration constants here
are chosen in such a way that $\rho$ is equal to zero at spatial infinity,
so that $\rho$ does not change instantaneously outside of the collision
region either.

The initial and final configurations of the fields are shown in figs. 1 and 2,
respectively, by solid lines;
the final configuration
 of the conventional proccess with energy conservation
is shown  by dashed lines in fig. 2 for comparison.  
In fig. 2 and in what follows we consider for definiteness the case
\[
   C>0\;,~~~~~\mu_i > \mu_f
\]
The process shown in these figures is perfectly local and causal:
the deviations of all fields from their conventional values
propagate out of the collision region (where energy is assumed to
be violated) with the speed of light.
The final dilaton 
field $\phi$ is the same as that of   the conventional process;
in particular, its long-range behavior  is not affected by 
energy non-conservation. On the other hand, the field $\rho$ in the final 
state is non-trivial and corresponds to longitudinal gravitational waves.
It is the presence of these longitudinal waves that ensures the validity of
the constraints after the collision, even  though the energy 
of matter is not conserved and the dilaton field does not change
asymptotically. 

\begin{figure}[htb]
\begin{center}
\epsfig{file=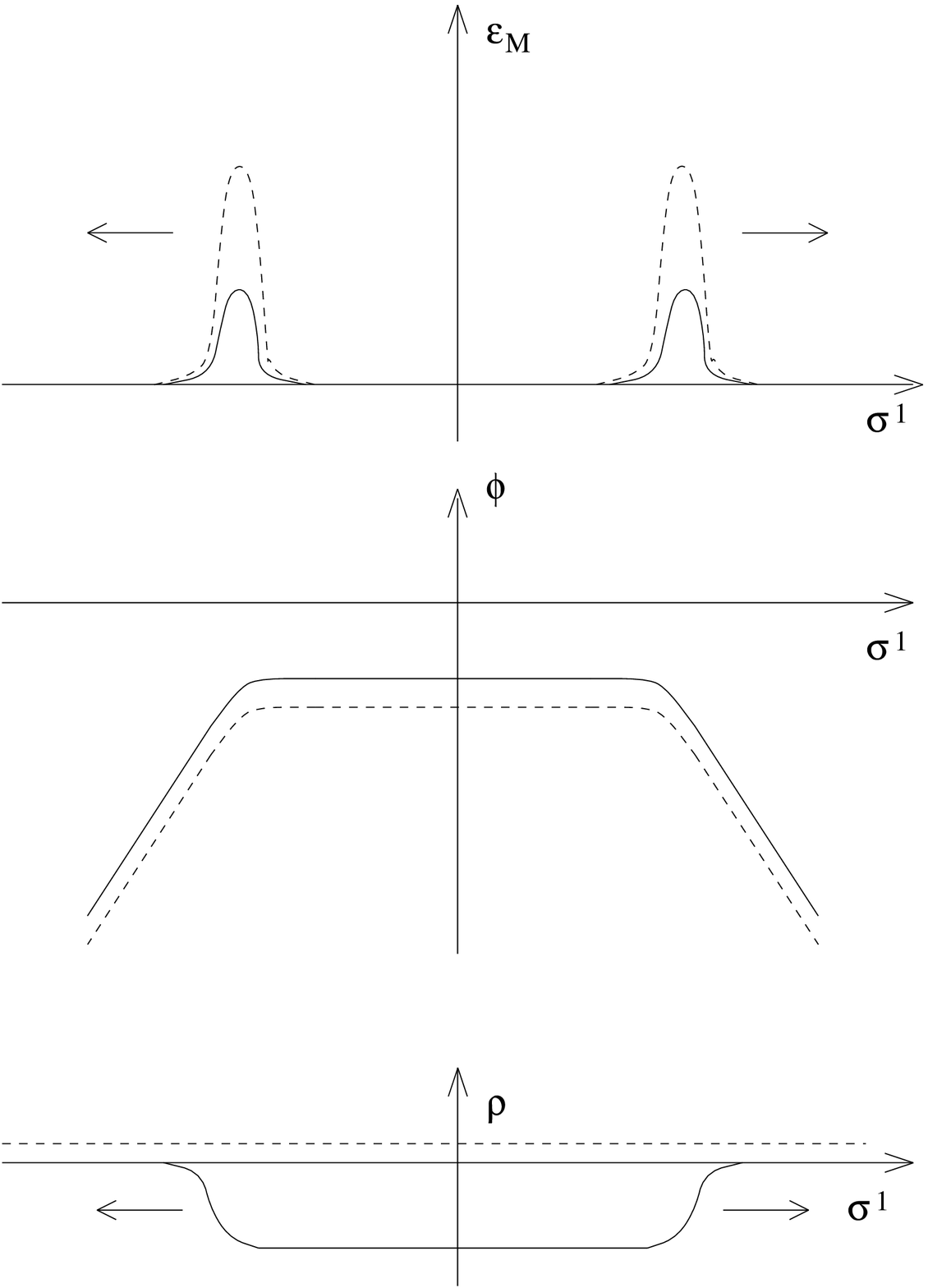,width=4in,height=4.5in}
  Fig.~2.
\end{center}
\end{figure}

Of course, the longitudinal $\rho$-wave of the final state
may be gauged away.
The corresponding gauge transformation is non-trivial far away from
the collision region, i.e., this gauge transformation describes
the change of the reference frame everywhere, including spatial infinity.
As this gauge transformation provides additional insight, let us
perform it explicitly. In the limit of infinitely narrow pulses the gauge
transformation to the frame where $\rho'(\sigma')=0$ is determined by the
following relations between new and old coordinates $\sigma'$ and $\sigma$,
\[
    \log \frac{d\sigma_{+}'}{d\sigma_{+}} = 
      -\frac{1}{C}(\mu_i - \mu_f) \epsilon (\sigma_{+})
\]
\[
    \log \frac{d\sigma_{-}'}{d\sigma_{-}} = 
      -\frac{1}{C}(\mu_i - \mu_f) \epsilon (\sigma_{-})
\]
At {\it large positive} $\sigma_1$ (i.e., to the right of the pulses) we
have
\[
    \sigma_{+}' = \e^{\eta} \sigma_{+} + \const\;,~~~~~
    \sigma_{-}' = \e^{-\eta} \sigma_{-} + \const
\]
with
\[
   \eta =  -\frac{1}{C}(\mu_i - \mu_f)
\]
This correspond to a Lorentz boost in the negative direction. In other words,
an observer, that is at rest in the old coordinate system, moves 
towards the pulses  with 
rapidity $\eta$ in the
frame where space-time is flat. This explains
why the long-ranged dilaton field measured by this observer remains 
equal to the 
initial one even though energy is not conserved.

At {\it large negative} $\sigma_1$ we have
\[
    \sigma_{+}' = \e^{-\eta} \sigma_{+} + \const\;,~~~~~
    \sigma_{-}' = \e^{\eta} \sigma_{-} + \const
\]
so the original observer on the left of the pulses also moves 
towards the pulses, as viewed from the new frame. Thus, the inertial
observers
on the left and on the right of the pulses, that were at rest
before the collision, move towards each other after the collision
(if energy is not conserved). To establish this fact, however,
the two observers have to communicate with each other, so they are
able to find out that the total energy of the pulses has changed
only after they exchange signals\footnote{Note that the definition
(\ref{9c}) of the ADM mass is given in flat space-time. Note also
that this definition involves the dilaton field both to the left and to
the right of the pulses.}. This explains the causal nature of the
whole process.

Thus, we find that energy non-conservation is compatible with locality
and causality in (1+1) dimensions even in the presence of long-ranged
fields of Newtonian type. This possibility may well be peculiar to (1+1)
dimensions, as only in that case  spatial infinity is disconnected.
Also,  Birkhoff's theorem does not hold 
 in (1+1) dimensions, which seems
to be of importance too. Still, it remains to be understood whether similar
possibility of local and causal energy non-conservation exists in
higher dimensions.

The author is indebted to A.A. Tseytlin for helpful correspondence
and to P.G. Tinyakov for useful discussions. This work is supported in 
part by the
Russian Foundation for Basic Research, project 96-02-17449a, by INTAS 
grant 93-1630-ext and CRDF grant 649.

\end{document}